\documentclass[12pt,preprint]{aastex}

\newcommand{\xte}{{\sl RXTE\/}}

\newcommand{\teff}{T$_{\rm eff}$}

\newcommand{\eps}{erg\,s$^{-1}$}
\newcommand{\epcs}{erg\,cm$^{-2}$s$^{-1}$}

\newcommand{\simpropto}{\lower.2ex\hbox{$\; \buildrel \sim  \over \propto \;$}}

\shorttitle{The Fall and the Rise of X-rays from Dwarf Novae}
\shortauthors{Fertig et al.}

\begin{document}


\title{The Fall and the Rise of X-rays from Dwarf Novae in Outburst:
       \xte\ Observations of VW Hydri and WW Ceti}


\author{D. Fertig\altaffilmark{1}, K. Mukai\altaffilmark{2},
        T. Nelson\altaffilmark{3,4}, and J.K. Cannizzo\altaffilmark{5}}
\affil{Department of Physics, University of Maryland,
       Baltimore County, 1000 Hilltop Circle, Baltimore, MD 21250.}

\altaffiltext{1}{Present address: Department of Physics \& Astronomy,
                 George Mason University, MSN 3F3, Fairfax, VA 22030}
\altaffiltext{2}{Also CRESST and X-ray Astrophysics Laboratory, NASA/GSFC,
                 Greenbelt, MD 20771}
\altaffiltext{3}{Also X-ray Astrophysics Laboratory, NASA/GSFC,
                 Greenbelt, MD 20771}
\altaffiltext{4}{Present address: School of Physics and Astronomy,
                 University of Minnesota, 116 Church St. SE, Minneapolis,
                 MN 55455}
\altaffiltext{5}{Also CRESST and Astroparticle Physics Laboratory, NASA/GSFC,
                 Greenbelt, MD 20771}



\begin{abstract}
In a dwarf nova, the accretion disk around the white dwarf is a source of
ultraviolet, optical, and infrared photons, but is never hot enough
to emit X-rays.  Observed X-rays instead originate from the boundary
layer between the disk and the white dwarf.  As the disk switches
between quiescence and outburst states, the 2--10 keV X-ray flux is usually
seen to be anti-correlated with the optical brightness.  Here
we present \xte\ monitoring observations of two dwarf novae, VW Hyi and
WW Cet, confirming the optical/X-ray anti-correlation in these two systems.
However, we do not detect any episodes of increased hard X-ray flux on the
rise (out of two possible chances for WW Cet) or the decline (two for
WW Cet and one for VW Hyi) from outburst, attributes that are clearly
established in SS Cyg.  The addition of these data to the existing
literature establishes the fact that the behavior of SS Cyg is the
exception, rather than the archetype as is often assumed.  We speculate
on the origin of the diversity of behaviors exhibited by dwarf novae,
focusing on the role played by the white dwarf mass.
\end{abstract}




\section{Introduction}

The outburst cycles of dwarf novae are understood as due to thermal
instability in the accretion disk around the white dwarf primary.
In the simplest form of the theory, the disk traces an S-curve in
the surface density ($\Sigma$)--effective temperature (\teff) plane,
with two stable branches.  The mass transfer rate from the Roche-lobe
filling secondary is sufficiently low in dwarf novae that the disk is
usually in the quiescent branch in which \teff\ is low and hydrogen
is mostly neutral.  The disk is completely out of steady state, with
accretion rate $\dot m$ at each radius $r$ varying as
$\dot m(r)$ \simpropto $r^3$, so the accretion rate onto the white dwarf
is low and $\Sigma$ increases over time.  When it reaches the end of the
quiescent branch, the disk jumps to the outburst branch in which the
disk is hot, ionized, and efficiently transports matter onto the
white dwarf.  See \cite{L2001} for a comprehensive review of the
disk instability theory.

Dwarf novae are well known sources of 2--10 keV X-rays, particularly
during quiescence but also in outburst.  Since the accretion disk 
surrounding a white dwarf is never hot enough to generate X-rays,
even during outburst, we must seek their origin elsewhere.
A widely held view is that they originate in an equatorial boundary
\citep{PR1985a,PR1985b}, which connects the Keplerian accretion disk
to the slowly rotating white dwarf surface.  The detection of a narrow
X-ray eclipse in OY Car (e.g., \citealt{WW2003}) and other deeply
eclipsing dwarf novae in quiescence is consistent with this picture,
while casting doubt on the ``coronal siphon flow'' model of \cite{MM1994},
which posits a hot, spatially extended X-ray emitting region around
the white dwarf.  In the rest of this paper, we therefore proceed with
the boundary layer model as the basis of our interpretation.

In this view, the quiescent boundary layer is optically thin and hot
($kT \sim$10 keV) while the outburst boundary layer is predominantly
optically thick with ($kT\sim$10 eV) but also contains an optically
thin region.  Coordinated multi-wavelength observations of dwarf novae
can therefore reveal the physical states of the accretion disk and the
boundary layer, at different points during the outburst cycle.
Previous campaigns during a single outburst include those on SS Cyg
\citep{WMM2003}, VW Hyi \citep{Wea1996,Hea1999}, and GW Lib \citep{Bea2009}.
These three campaigns in particular had dense enough X-ray coverage through
the outburst to enable the authors to study the time evolution of X-ray
flux.  Campaigns that covered multiple outburst cycles include those on
YZ Cnc \citep{VWM1999}, V1159 Ori \citep{Sea1999}, and SU UMa \citep{CW2010}.
Here we present our analysis of two sets of X-ray observations taken with
the PCA instrument \citep{PCA} on board \xte\ \citep{RXTE}, one of
a superoutburst of VW Hyi, the other of WW Cet obtained over three months
and containing two outbursts.  These data sets are well suited to the
study of transitions of the X-ray emissions from quiescence to outburst
and back.

\section{Observations and Data Reduction}

VW Hyi is a dwarf nova with a quiescent magnitude of V$\sim$14 and
an orbital period of 1.78 hr \citep{SV1981}.  In addition to the
normal outbursts (once every 20--30 d), it has superoutbursts
once every 180 d or so, the defining characteristic of the SU UMa
subclass.  The superoutbursts are brighter (V$\sim$8.5 vs. 9.5) and
last longer (10--14 d vs. 3--5 d), and are characterized by
superhumps, which are photometric variations at a period slightly
longer than the orbital period ($\sim$1.84 hr).  \cite{Sea2006}
estimated its white dwarf to be 0.71$^{+0.18}_{-0.26}$ M$_\odot$
based on a refined measurement of the gravitational redshift in the
photospheric absorption line. \cite{Pea1987} contains further details
of VW Hyi.

VW Hyi was observed 42 times with \xte\ between 1996 May 02 13:08 UT and
1996 May 16 23:55 UT (obsIDs 10041-02-01-00 through 10041-02-44-00
except 10041-02-31-00 and 10041-02-36-00), with a typical
exposure of 2 ks each and separated by 4--16 hr.  According
to the AAVSO data base, VW Hyi went into an outburst around
JD 2450201.0 (1996 Apr 27 at 12 UT), brightening from magnitude
13.5 to $\sim$9 in about 6 hr, and returned to quiescence (V$>$13)
at about 2450216.25 (May 12 at 18 UT).  Both the peak magnitude
(V$<$9) and the duration establish this as a superoutburst,
even though we do not have direct confirmation through the
detection of superhumps.  Thus, the \xte\ observations started
$\sim$5 d after the start of a superoutburst of VW Hyi
and continued $\sim$4 d after its return to quiescence.

WW Cet is a dwarf nova with a 4.22 hr orbital period, with an average
quiescent magnitude of V=13.9 and reaching V$\sim$11 in outburst,
and also showing occasional low states as faint as V=16.2 \citep{Rea1996}.
\cite{Hea1990} carried out radial velocity measurements of the accretion
disk and the secondary, and assuming a main-sequence companion, estimated
the white dwarf mass to be 0.83$\pm$0.16 M$_\odot$ for this system.
Recently, \cite{SS2011} have detected a standstill in WW Cet for
the first time, establishing this to be a Z Cam type dwarf nova.  However,
at the time of our observations, it displayed the normal quiescence-outburst
behavior of dwarf novae.

WW Cet was observed every day for 90 consecutive days between 2002
Aug 14 and 2002 Nov 11 (obsIDs 70005-01-01-00 through 70005-01-90-00),
with a typical daily exposure of 2.5--3.5 ks.  On 9 occasions, the
observations were split into two sub-exposures (Oct 25, 26, 27, 29;
Nov 1, 2, 3, 10, and 11), and once (Nov 8) into three sub-exposures,
which were correspondingly shorter.  Treating them separately,
we therefore have 101 observations of WW Cet over a 3-month period.
According to the AAVSO data base, WW Cet went into outburst twice
during this campaign; we will describe them in more details below.
The last outburst before 2002 Aug 14 was poorly covered but appears
to have ended on or around Jul 12; the first outburst after Nov 11
started on Dec 10.

We analyzed the standard 2 binned mode data (i.e., 129 channel spectra
accumulated every 16 s) using HEAsoft v6.6.3.  We screened the data
using the following screening criteria: angular separation between
target and the pointing direction of $<0.05^\circ$, elevation angle
$>5^\circ$, time since the last SAA passage of $>$25 min, and electron
contamination low (ELECTRON2$<$0.1).  PCA background was estimated
using L7 faint model.  VW Hyi observations were taken with 3 to 5 of
the 5 PCUs active. Observations of WW Cet, on the other hand, were obtained
with 2 or 3 active PCUs.  This, however, included PCU0, which by
2002 had lost its propane layer, leading to a larger uncertainty in
the background modeling.  We therefore excluded PCU0 data from further
analysis for WW Cet (but not for VW Hyi).

Since the PCA has little sensitivity below 2.5 keV, and these sources
are not significantly detected above 10 keV, we analyze the background
subtracted date in the 2.5--10 keV range.  Since our main interest is
the variability through an outburst or through an interoutburst period,
our main data products are the average count rates per active PCU per
pointing.    We also performed spectral fitting for completeness,
even though the data quality is rather limited for this purpose.

\section{Results}

\subsection{VW Hyi}

\begin{figure}
\plotone{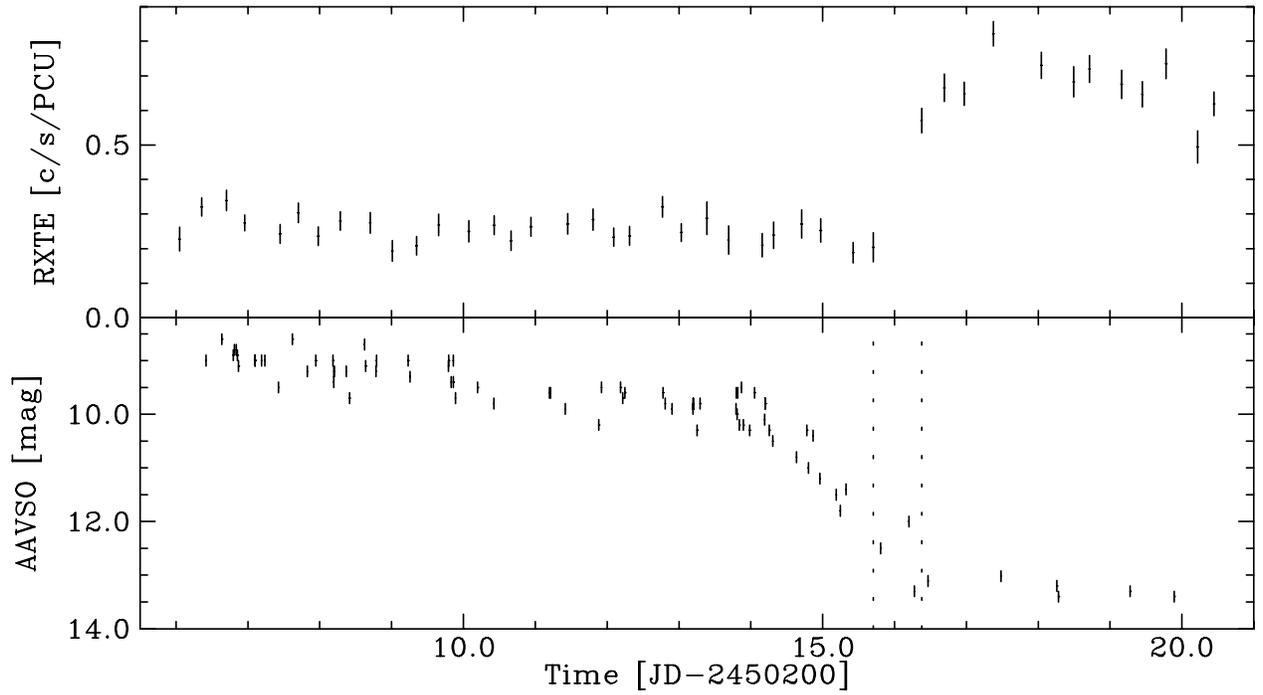}
\caption{The X-ray (top panel) and visual (bottom) light curves of
VW Hyi during 1996 May.  The X-ray data were obtained with \xte\ PCA
while the optical data were collected by amateur observers and compiled
by AAVSO.  Two vertical dashed lines mark the time of the last ``outburst''
and the first ``quiescent'' X-ray points (see text).}
\label{vwov}
\end{figure}

We present the X-ray and visual light curves of VW Hyi in
Figure\,\ref{vwov}.  There is a clear dichotomy in the observed
\xte\ count rates: the first 30 points (taken during the 10 day
period covering the superoutburst peak and decline) are at
0.25$\pm$0.04 c\,s$^{-1}$PCU$^{-1}$ (mean and standard deviation).
The last 12 points, taken after the return to quiescence, are at
0.67$\pm$0.08 c\,s$^{-1}$PCU$^{-1}$.

We have fitted the average quiescent spectra of VW Hyi using a
single temperature {\tt mekal} model \citep{Mea1985,Mea1986,Lea1995,Kea1996}
with a Gaussian emission line at 6.4 keV and a neutral absorber.
This gives an acceptable fit ($\chi^2_\nu$=0.74) with $kT=6.4\pm$1.3 keV,
metallicity 0.44$\pm$0.22 solar, and a 240 eV equivalent width line at 6.4 keV,
and a 2--10 keV flux of 7.3$\times 10^{-12}$ \epcs.  Although
multi-temperature models are more physical and known to be required in
fitting higher-quality X-ray data on VW Hyi \citep{Pea2005} and result
in even lower $\chi^2_\nu$ with the \xte\ data, they are not required
due to the modest spectral resolution of the PCA.

We have also attempted spectral fits to the average superoutburst
spectrum of VW Hyi for completeness.  Although our analysis indicates
a nominal 2--10 keV flux of 2.7$\times 10^{-12}$ \epcs, this is below
the confusion limit for the PCA data of
4$\times 10^{-12}$ \epcs\ \citep{PCACAL} set by the Poissonian fluctuation
in the cosmic X-ray background within the PCA field-of-view in the direction
of VW Hyi.  Therefore, while we can be confident that VW Hyi was fainter
in superoutburst by 0.42$\pm$0.09 c\,s$^{-1}$PCU$^{-1}$ (approximately
5$\times 10^{-12}$ \epcs) than in quiescence, we cannot determine
the actual flux level or the spectral shape during superoutburst from
our data.

The X-ray state transition took place in 17 hr.
The last superoutburst obsID is 10041-02-30-00, with good exposure
between 05:09 and 05:37 UT on 1996 May 12.  The average 2.5--10 keV
count rate is 0.20 c\,s$^{-1}$ and there is no significant change in brightness
within the exposure.  The first quiescence obsID is
10041-02-32-00\footnote{It is likely that an obsID 10041-02-31-00 was
planned between these exposures, but was canceled due to a higher priority
observation.}, which started 21:10 UT on the same day, by which time
it was already at 0.57 c\,s$^{-1}$.  During this interval, VW Hyi
declined from V$\sim$12.5 (at 07:24 UT) to $\sim$13.3 (at 18:37 UT),
close to the quiescent level.

Our best-fit quiescent model predicts an {\sl EXOSAT\/} ME rate of
0.6 c\,s$^{-1}$ per half array, which is in reasonable agreement with
the observations of \cite{vdWH1987}.  It also predicts $\sim$3
c\,s$^{-1}$ with {\sl Ginga\/} LAC, somewhat lower than the observed
rate of $\sim$5 c\,s$^{-1}$ of \citep{Wea1996} . This {\sl Ginga\/}
observation also included the outburst to quiescence transition.
In this case, the strong variability within each orbit makes it harder
to judge the exact duration of transition (see their Figure 2), but it was
definitely shorter than 0.4 day, and was arguably as short as 0.1 day.
In the latter interpretation, the last $\sim$0.3 day of {\sl Ginga\/} data
may be interpreted as showing a slight rise from the initial quiescent
level of $\sim$4 c\,s$^{-1}$ to $\sim$7 c\,s$^{-1}$ during some
orbits.

\subsection{WW Cet}

\begin{figure}
\plotone{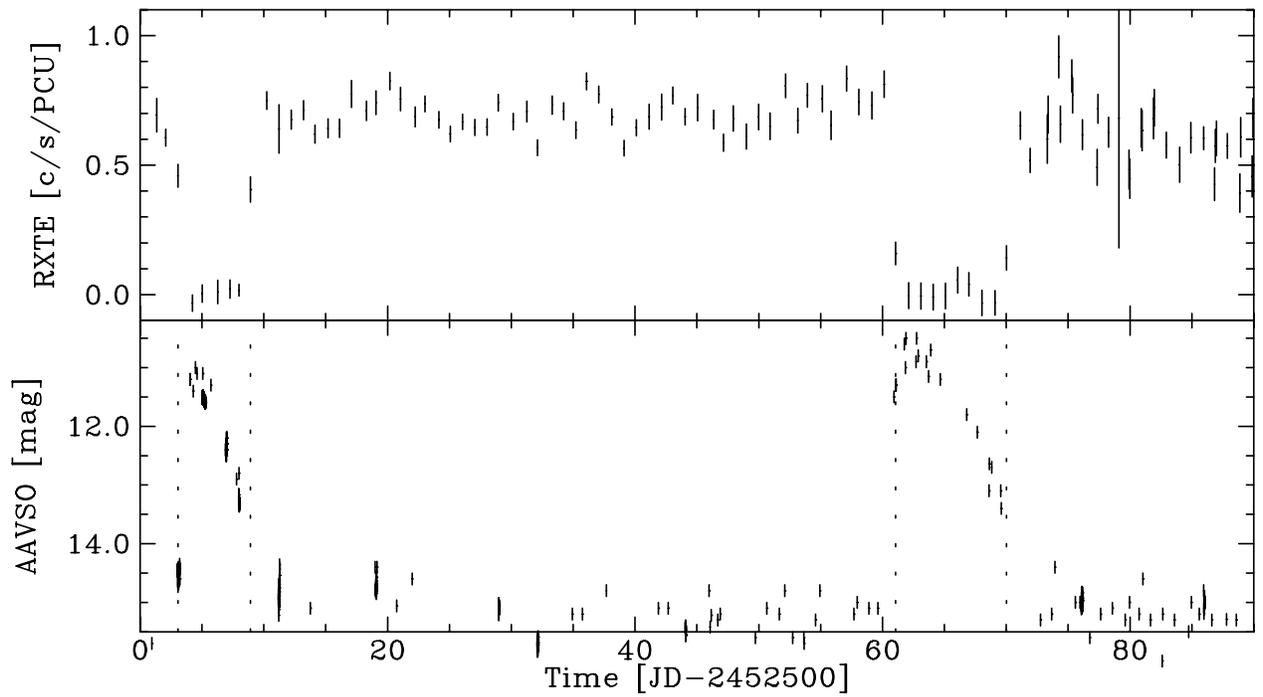}
\caption{The X-ray (top panel) and visual (bottom) light curves of
WW Cet during 2002 Aug--Nov.  The X-ray data were obtained with \xte\ PCA
while the optical data were collected by amateur observers and compiled
by AAVSO.}
\label{wwov}
\end{figure}

We present the X-ray and visual light curves of WW Cet in
Figure\,\ref{wwov}.  Similarly to the case of VW Hyi, WW Cet
shows a significant decrease in the X-ray flux level during
the two outbursts.  The average quiescent count rates and
standard deviations are 0.65$\pm$0.05 on Aug 14 \& 15,
0.70$\pm$0.07 during Aug 23--Oct 12, and 0.61$\pm$0.12 during Aug 26--Nov 11.
During outburst, the background-subtracted count rates are consistent
with zero (0.00$\pm$0.02 on Aug 17--21 and 0.00$\pm$0.03), well within
the systematic uncertainty in the background level.

The quiescent spectrum of WW Cet can be described using a single
temperature model ($\chi^2_\nu$=0.82) with $kT=7.0\pm$0.5 keV,
metallicity 0.50$\pm$0.10 solar, and a 67 eV equivalent width line
at 6.4 keV, and a 2--10 keV flux of 8.2$\times 10^{-12}$ \epcs.
The quiescent flux is clearly not constant, but there is a
sufficiently high level of point-to-point variability that we cannot
comment on any systematic trends.   Most importantly, we have fitted
a constant and a linear line to the fully covered interoutburst period
between Aug 23 and Oct 12.  We obtain a lower $\chi^2_\nu$ with the
constant model: That is, we do not detect a secular trend in the
quiescent X-ray flux of WW Cet.

We appear to have resolved the X-ray transitions into and out of
outbursts in WW Cet.  The count rates and statistical errors
on Aug 16, Aug 22, Oct 13, and Oct 22 are 0.46$\pm$0.04, 0.41$\pm$0.04,
0.16$\pm$0.04, and 0.14$\pm$0.04, intermediate between the quiescent
and outburst levels.  We therefore investigate the relative timing
of X-ray and optical transitions in more detail.

The Aug 16 X-ray data were taken between 13:17 and 14:11 UT.  In the optical,
AAVSO reports visible data on Aug 16 at 16:22 UT (V=14.6) and on Aug 17
at 12:30 UT (V=11.2).  Thus, it appears that X-ray decline began before a
significant rise in the optical.   The intermediate X-ray point at the end
of this outburst was taken on Aug 22 between 10:09 and 10:54 UT, which
unfortunately falls in a significant gap in the optical coverage (between
Aug 21 12:34 UT, when WW Cet was at V=13.34, and Aug 24 16:13 UT at V=14.96),
but was in any case close to the return to quiescence.
The Oct 13 X-ray data were taken between 13:27 and 14:21 UT.  In the optical,
WW Cet was already at V=11.5 at 10:15 UT.  The Oct 25 X-ray data were
taken between 20:00 and 20:13 UT, which is after it was seen at V=15.3
at 06:24 UT, therefore it appears that the X-ray recovery took place
only after the system had already returned to quiescence.

\subsection{Comparisons with Past Campaigns}

In terms of the boundary layer behavior, the single best observed outburst
of any dwarf novae may well be the 1996 October outburst of SS Cyg
\citep{WMM2003}.  This outburst had the following features:
(1) During the rise of the outburst, the $>$2 (``hard'') keV flux initially
increases (``enhancement -- rise''); (2) The optically thick boundary
layer develops, as evidenced by the sudden increase in the EUV flux
and a simultaneous drop in the hard X-ray flux is delayed relative to
the optical rise (``X-ray delay''); (3) Through the bulk of the outburst,
the hard X-ray flux is lower than in quiescence (``X-ray suppression'');
(4) During this period, the X-ray spectrum is softer than in quiescence
(``Spectral softening''); (5) The optically thick boundary layer disappears
near the end of the optical decay, as evidenced by decreased EUV flux
and increased hard X-ray flux (``Recovery at late decline''); and
(6) The hard X-ray flux remains elevated for a period after the recovery
(``Enhancement -- decline'').

It is worth examining the body of data accumulated to date to determine
if these features should be taken as the paradigm in which to interpret
observations of outbursts of other dwarf novae, for which only lesser
quantity and/or quality of data are available.  We have therefore
searched the literature for descriptions of well-observed outbursts,
and judged how many of these features are shared by other outbursts.

SU UMa shares 4 of the 6 characteristics with SS Cyg according to
\cite{CW2010}, but failed to show X-ray enhancements at rise or
in decline.  In the case of VW Hyi, the X-rays are suppressed during
outburst and recover at late decline.  However, the \xte\ campaign
did not cover the rise, and the {\sl ROSAT\/} data of \cite{Wea1996}
suggest little or no X-ray delay.  \cite{Hea1999} observed spectral
softening in VW Hyi in outburst in their {\sl BeppoSAX\/} data.
However, elevated X-ray flux levels have not been definitively detected
at rise or late decline, although the {\sl Ginga\/} LAC data allow the
possibility of enhancement at late decline.  As for WW Cet, we arrive
at the same set of answers as for VW Hyi based on the \xte\ campaign.
As for U Gem, the X-rays are in fact enhanced during outburst \citep{MMW2000},
and the outburst spectral shape is complex enough to preclude a simple
answer to the ``softening'' question \citep{Gea2006}.  GW Lib is another
dwarf nova with a higher X-ray flux during outburst \citep{Bea2009},
which was also harder.  We summarize these results in
Table\,\ref{tab:paradigm}.

In addition, \cite{Sea1999} contains an intriguing report that the
hard X-ray flux may have recovered during the middle of two superoutbursts
of V1159 Ori.  However, the quality of the non-imaging \xte\ data of
V1159 Ori is rather low, so we have not included this object in our
table.  Similarly, while \cite{VWM1999} shows that the X-ray fluxes
are suppressed during outburst in YZ Cnc, this campaign is less
conclusive regarding other questions.

\section{Discussion and Conclusions}

This paper adds to the literature of X-ray observation of dwarf novae
during outbursts.  Through these observations and analyses, common, if
not universal, patterns are emerging.

In the majority of dwarf novae, X-ray flux is suppressed during outburst.
This is true of 4 out of 6 systems listed in Table\,\ref{tab:paradigm},
probably also true in YZ Cnc and (largely) in V1159 Ori, as well as other
dwarf novae for which one or two X-ray spectra have been obtained in
outburst (e.g., Z Cam; see \citealt{BWO2001}).  We interpret
this as the consequence of an optically thick boundary layer, which
radiates most of the boundary layer luminosity during outburst.
In these majority, the X-ray spectrum also appears to become softer
in outburst.  We propose Compton cooling as a possible mechanism for this,
as \cite{Nea2011} postulated for the quiescent X-ray emission from RS Oph.

Standard textbooks such as \cite{FKR2002} convincingly show that
Compton cooling cannot be important in cataclysmic variables if the seed
photons are from the heated surface of the white dwarf in a 1-zone accretion.
However, the boundary layer in dwarf novae in outburst has at least two zones.
At the equator, the boundary layer is optically thick, emitting copious
soft photons; at somewhat higher latitude, the boundary layer is optically
thin but subject to these external seed photons (see Figure 8 of
\citealt{PR1985a}).  Since the number of soft photons greatly exceeds
that of the hot electrons (e.g., by a factor of 1,000 if the energy of a
10 keV electron is radiated away by 10 eV photons), each electron can
experience many Compton scattering events even though each photon scatters
once at most, thus keeping the soft spectrum largely unchanged (i.e.,
no hard inverse Compton component).  Since the density of a cooling flow plasma
rapidly increases towards lower temperatures, the optically thin region
of the boundary layer is predominantly Compton cooled at high temperatures
and predominantly Bremsstrahlung cooled at low temperatures.  The observed
soft-to-hard X-ray luminosity ratio for SS Cyg in outburst is $\sim 40$
\citep{WMM2003}, which makes it plausible that the optical thin part of
the boundary layer is partly Compton-cooled, since the two emitting regions
lie close to each other within the boundary layer.

To interpret the X-ray transitions, we must consider the disk
instability model in more detail.  The instability is principally
a local property of the disk.  However, once a transition from
the quiescent state to the outburst state happens in one part
of the disk, this propagates in the form of a heating front,
behind which the mass inflow rate becomes much higher.
X-ray behaviors are altered when the heating front reaches the
inner edge of the disk.   The optical flux, on the other hand,
predominantly arises from the outer parts of the disk, which have
the most emitting area.  It is likely that the outburst can start
at a variety of radial distances from the white dwarf in different
systems or in different outbursts, and this is described as
``outside-in'' and ``inside-out'' outbursts in the studies of
UV delay (see, e.g., \citealt{CWP1986}).  It is therefore possible
to have a range of X-ray delay times among systems, and among
different outbursts for a single dwarf nova. On the return to
quiescence, the cooling front is likely to start from near the
outer edge of the disk, and so X-ray transition happens when it
reaches the boundary layer, near the end of optical decline.

From our own analysis and a survey of the literature, we conclude
that the temporary enhancement of X-ray flux during the rise and late
decline has only been definitively seen in SS Cyg so far.  There
have only been two studies that attempt to incorporate the
observational constraints imposed by the X-ray observations
of dwarf novae throughout the course of an outburst  into the disk
instability model: The calculations of \cite{SHL2003} were specifically
for SS Cyg, while those of \cite{SHL2004} were tailored for VW Hyi.
These papers adopt a critical value of
${\dot M}_{\rm crit} = 10^{16}$ g s$^{-1}$ in the inner disk as the dividing
line between having an optically thin or thick boundary layer, following
the empirical determination by \cite{PR1985b}.  A comparison of the
theoretical X-ray fluxes shown in Fig. 4 of \cite{SHL2003} with
those of Fig. 9 of \cite{SHL2004} shows the X-ray enhancements, both
during rise and decline, to be common to both models.
Thus, the model can account for the behavior exhibited by SS Cyg
(see Fig. 2 of \citealt{WMM2003}), but not those of SU UMa, VW Hyi,
or WW Cet (Table 1).

\begin{figure}
\plotone{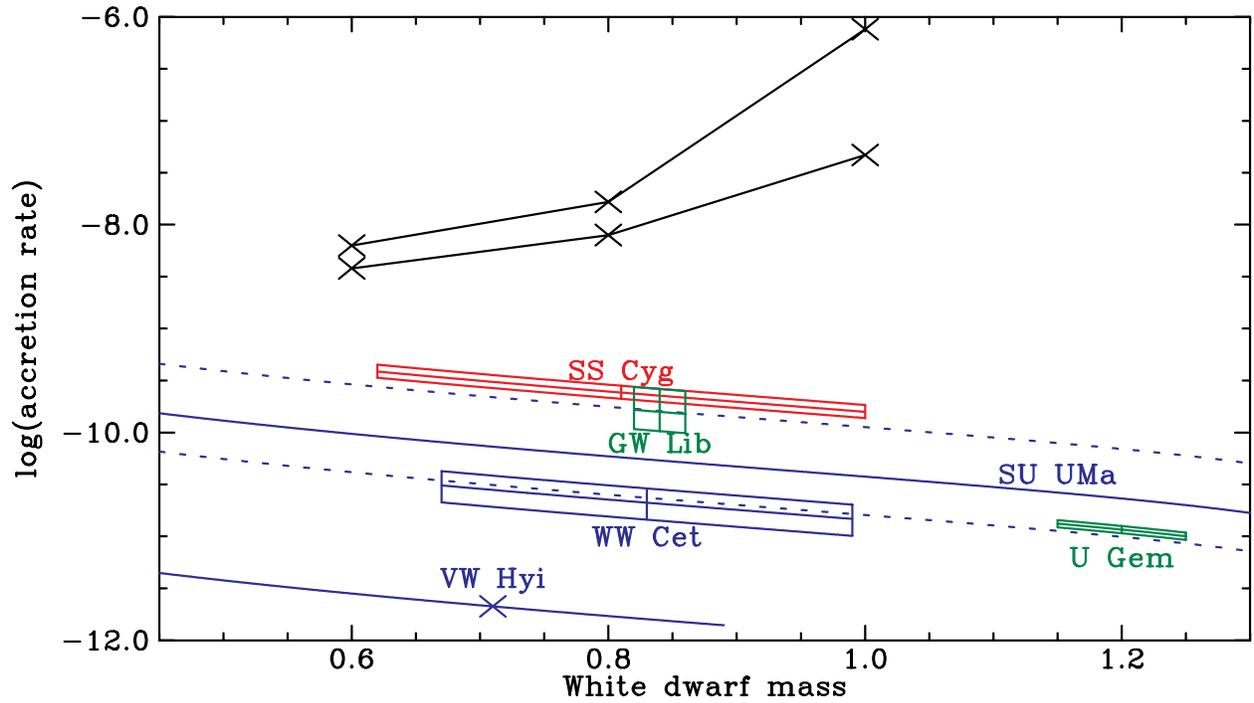}
\caption{The inferred accretion rate at the highest flux of the optically
thin X-ray emission, plotted against the white dwarf mass (see text for
object-by-object details).  Also shown are the theoretical curves of
optically thin-to-thick transition from \cite{PN1995}.}
\label{mcrit}
\end{figure}

To investigate this further, we have derived the critical accretion rate
observationally for the 6 systems listed in Table\,\ref{tab:paradigm}.
We assume that the highest luminosity observed for the optically thin
component reflects (or is close to) this critical value.  We have collected
the information in Table\,\ref{tab:mcrit}, including the adopted values and
references for the distances and the white dwarf mass.  We convert the
bolometric X-ray luminosity to accretion rate using $L=GM_1 \dot m/2R$,
for a range of plausible values of M$_1$.  We also plot the results in
Figure\,\ref{mcrit} against M$_1$.  For five systems, the plots are
limited to the range of M$_1$ inferred from the reference cited.
For SU UMa, we show three parallel lines over the entire mass range
of the figure, as there are no published estimates for M$_1$.
The objects are color-coded based on the presence or otherwise of
hard X-ray suppression and temporary rise.  For comparison, we also plot
the theoretical curves from Figure\,8(a) of \cite{PN1995}.

Unfortunately, we are unable to reach a firm conclusion, except
that the observationally derived values are all much lower than what
\cite{PN1995} predicted.  This may not be a huge problem, since the
specific values depend on the assumptions made in the study, including
the adopted value for the viscosity parameter, $\alpha$.  The authors
themselves caution against adopting these precise theoretical values,
although the strong dependence on M$_1$ is likely to be a more robust
conclusion.  The values we derived observationally are in closer
agreement with the empirical conclusion of \cite{PR1985b}.

Nevertheless, we propose that the diversity of behavior shown by these 6
dwarf novae reflects the fact that ${\dot M}_{\rm crit}$ is in fact a
function of various system parameters, and not uniform among all dwarf novae.
In this view, the quiescent accretion rate is already close to this value
in some systems, in which case any increase during outburst will
immediately lead to the development of an optically thick boundary
layer and hard X-ray suppression.  Moreover, we propose that the primary
mass is likely to be a key parameter, partly because the conversion from
luminosity to accretion rate depends on $M_1$, and partly because the
theoretical investigation by \cite{PN1995} indicates that the critical
rate depends steeply on M$_1$.

Looking at individual systems, the inferred critical accretion rate for
VW Hyi is extremely low, compared with those for other systems as well
as theoretical expectations.  We note, however, that the distance estimate
for this system relies on an indirect argument without any indication
of the size of the error bar, and should be refined.    As for the two
systems that do not show X-ray suppression during outburst, GW Lib can
be understood as due to the extremely low accretion rate in quiescence
\citep{Hea2007}.  Interestingly, the inferred accretion rate for U Gem
is also low.  If the current mass estimate is reliable, it also has the
most massive white dwarf of the 6 systems, which suggests that the
critical accretion rate is also highest.  However, both these systems
are reported to have developed an optically thick boundary layer, so
the lack of hard X-ray suppression is a puzzle, particularly in light
of our Compton-cooling interpretation for systems that do show hard
X-ray suppression.

We believe that we should no longer assume a single critical accretion
rate for all dwarf novae.  Rather, we propose that future efforts be
focused on determining which system parameters determine the critical
accretion rate.  Specifically, X-ray observations of more dwarf
novae with well-determined distances and white dwarf masses should be
obtained through outburst cycles.  In particular, systems with extreme
masses should be high priority targets, given the likely steep dependence
of the critical accretion rate on the white dwarf mass.

\acknowledgments

This research has made use of data obtained from the High Energy Astrophysics
Science Archive Research Center (HEASARC), provided by NASA's Goddard Space
Flight Center.  We acknowledge the planning effort by Drs. E. Schlegel and
D. Baskill, who were the original PIs of VW Hyi and WW Cet observations,
respectively.  We also acknowledge with thanks the variable star observations
from the AAVSO International Database contributed by observers worldwide and
used in this research.

\clearpage

\begin{deluxetable}{lcccccc}
\rotate
\tablecaption{The SS Cygni Paradigm\label{tab:paradigm}}
\tablehead{\colhead{Object} & \colhead{X-ray} & \colhead{Spectral} &
           \colhead{X-ray} & \colhead{Recovery at} &
           \colhead{Enhancement} & \colhead{Enhancement} \\
            & \colhead{Suppression} & \colhead{Softening} & \colhead{Delay} &
           \colhead{late Decline} & \colhead{-- Rise} & \colhead{-- Decline} }

\startdata
SS Cyg    & Yes & Yes & Yes & Yes & Yes & Yes \\
SU UMa    & Yes & Yes & Yes & Yes & No  & No  \\
VW Hyi    & Yes & Yes & No? & Yes & No  & Maybe \\
WW Cet    & Yes & ?   & No? & Yes & No  & No  \\
U Gem     & No  & ?   & Yes & Yes & ?   & ?   \\
GW Lib    & No  & No  & ?   & ?   & ?   & ?   \\
\enddata
\end{deluxetable}

\begin{deluxetable}{lccccc}
\tablecaption{Critical Mass Accretion Rate\label{tab:mcrit}}
\tablehead{\colhead{Object} & \colhead{Distance\tablenotemark{a}} &
           \colhead{Primary Mass\tablenotemark{b}} &
           \colhead{L$_{X,peak}$\tablenotemark{c}}  & \colhead{$\dot m$} &
           \colhead{Fiducial M$_1$} \\
            & \colhead{(pc)} & \colhead{(M$_\odot$)} & \colhead{(\eps)} &
           \colhead{g\,s$^{-1}$} & \colhead{M$_\odot$} }

\startdata
SS Cyg    & 165$^{+13}_{-11}$  & 0.81$\pm$0.19          &
            1.2$\times 10^{33}$ & 1.0$\times 10^{16}$ & 1.0 \\
SU UMa    & 260$^{+190}_{-90}$ &                        &
            2.6$\times 10^{32}$ & 4.2$\times 10^{15}$ & 0.75 \\
VW Hyi    & 65                 & 0.71$^{+0.18}_{-0.26}$ &
            8.1 $\times 10^{30}$ & 1.3$\times 10^{14}$ & 0.71  \\
WW Cet    & 146$\pm$25         & 0.83$\pm$0.16          &
            1.1 $\times 10^{32}$ & 1.3$\times 10^{15}$ & 0.83  \\
U Gem     & 100$\pm$4          & 1.20$\pm$0.05          &
            1.4 $\times 10^{32}$ & 7.3$\times 10^{14}$ & 1.20  \\
GW Lib    & 104$^{+30}_{-20}$  & 0.84$\pm$0.02          &
            8.4$\times 10^{32}$ & 7.0 $\times 10^{15}$ & 1.0 \\
\enddata

\tablenotetext{a}{Distances are taken from \cite{Hea2004} for SS Cyg and U Gem,
                  \cite{T2003} for SU UMa and GW Lib, \cite{W1987} for VW Hyi,
                  and \cite{SHM1996} for WW Cet.}
\tablenotetext{b}{The white dwarf masses are taken from \cite{BRB2007} for SS
                  Cyg, \cite{Sea2006} for VW Hyi, \cite{Eea2007} for U Gem,
                  and \cite{vSea2010} for GW Lib.  There is no estimate in
		  the literature for SU UMa, while the numbers for WW Cet
                  uses the mass ratio obtained by \cite{Hea1990} and assumes
		  a main-sequence secondary.}
\tablenotetext{c}{The peak X-ray luminosities are taken from \cite{WMM2002}
                  for SS Cyg, \cite{CW2010} for SU UMa, \cite{Pea2005} for
		  VW Hyi and WW Cet, \cite{Gea2006} for U Gem and
		  \citep{Bea2010} for GW Lib.  These values are then used
                  by the same authors to infer the highest accretion rate
		  through the optically thin boundary layer for the fiducial
		  white dwarf mass.}
\end{deluxetable}

\clearpage


\end{document}